\documentclass[12pt]{article}
\usepackage{graphicx}
\newcommand{\vs}{\vspace}
\newcommand{\hs}{\hspace}

\newcommand{\be}{\begin{equation}}
\newcommand{\bea}{\begin{eqnarray}}
\newcommand{\ben}{\begin{eqnarray*}}
\newcommand{\pa}{\partial}

\newcommand{\de}{\delta}
\newcommand{\ba}{\begin{array}}
\newcommand{\ea}{\end{array}}

\newcommand{\ity}{\infty}

\newcommand{\lt}{\left}
\newcommand{\La}{\Lambda}

\newcommand{\ri}{\right}
\newcommand{\fr}{\frac}
\newcommand{\ep}{\epsilon}

\newcommand{\noi}{\noindent}

\newcommand{\bc}{\begin{center}}
\newcommand{\ec}{\end{center}}
\newcommand{\ee}{\end{equation}}
\newcommand{\eea}{\end{eqnarray}}
\newcommand{\een}{\end{eqnarray*}}

\newcommand{\al}{\alpha}
\newcommand{\ga}{\gamma}
\newcommand{\ra}{\rightarrow}
\newcommand{\Ga}{\Gamma}
\newcommand{\la}{\lambda}

\newcommand{\bt}{\beta}

\begin{document}

\baselineskip=6mm


\baselineskip=5mm
\vs{2mm}

\noi {\bf  POWER LAWS IN WALL AND WAKE LAYERS OF A TURBULENT \\ BOUNDARY LAYER\footnote{The paper was presented at $7^{th}$ Asian Congress of Fluid Mechanics  Dec 8-12, 1997, Chennai (Madras) pp. 805-808 (Editors: Hiroshi Sato, R. Narasimha,  Peiyuan Chou).}$^,$\footnote{
Egolf, P.W. and  Hutter, K. 2020 Nonlinear, Nonlocal and Fractional Turbulence,  Springer --- pp 65-67:

\noi Table 5.3 The preference for a logarithmic and a power law was alternating over the                    last hundred years:

\noi 1938 Millikan (1938) Similarity hypothesis favors the logarithmic law

\noi 2001 Afzal (2001) Shows that similarity consideration of Millikan can also be applied                                to confirm a power law.

In a publication of  Cipra (1996) in "Science" Chorin is cited: "The 'law of the wall' was viewed as one of the few certainties  in the difficult field of turbulence, and now it has been dethroned". He further continues by writing  "Generations of engineers, who learned the law, will have to abandon it."
({\em Cipra, B.  1996 A new theory of turbulence causes a stir among experts. Science. 272 (5264), page 951. https://doi.org/10.1126/science.272.5264.951}).

Then at the very beginning of the new millennium Afzal (2001) demonstrated that the Millikan/Clauser idea can also be applied to prove consistency with a power law as  averaged velocity profile of the overlap region. This work then immediately removed a believed theoretical obstacle.  
 ({\em Afzal, N. 2001 Power law and log law velocity profiles in turbulent boundary-layer flow: equivalent relations at large Reynolds numbers. 
Acta Mechanica 151, pp. 195 - 210}).
}
}

\vs{3mm}
\noi {\bf \small Noor Afzal\hs{10mm} (Email ID: noor.afzal@yahoo.com)\\ 
Faculty of Engineering, Aligarh Muslim University, Aligarh 202 002, India.
}

\noi {\bf ABSTRACT:} The power laws in the overlap region of two layer theory of turbulent 
boundary layer have been obtained by Izakson-Millikan-Kolmogorov hypothesis. The solution of the open functional equation is not unique. 
The power laws  for velocity \& skin friction have  equivalence with log laws  for velocity \& skin friction, for large Reynolds numbers.
\vs{2mm}

\noi {\bf 1. Introduction}\\
	Narasimha [1] proposed that Barenblatt [2] recent analysis of relation between power law and log law velocity profiles, reviving an issue raised already in Prandtl [3], call for accurate experimental data at high Reynolds number. Further, the pipe flow velocity profile exhibits a very week defect layer, so a more severe test for power law velocity profile would be high Reynolds number turbulent boundary layer(where the overlap would not constitute what Barenblatt calls "main body of flow").

	For pipe data, the power index $\al$ and multiplying coefficient $C$ in the power law velocity profile were determined as the function of Reynolds number $Re = U_s d/\nu $ (based on average speed $U_s$  and pipe diameter $d$), by Nikuradse and Weighardt (see Schlichting [4] pp. 563-565) whereas Barenblatt [2], Kailasnath [5], Afzal [6] and Zagarola , Perry \& Smits [7] proposed correlations for $\al $ and $C$. Analysis of Afzal [6] proposed
$$
\al = 1/ \ln \: R_\tau , \hs{10mm} C \exp (1) = (k \al)^{-1} + B, \hs{10mm} R_\tau = u_\tau \de / \nu
\eqno(1)
$$
where $k$ is Karman constant and $B$ is intercept of wall log law. Further, $u_\tau$ is friction velocity and $\de$ is the boundary layer thickness or pipe radius as appropriate.

	The analysis of renormalization theory shows (see Afzal \& Narasimha [8]) that for pipe flow $\al = 3/(2 \ln \: Re)$ proposed by Barenblatt [2] is not consistent  with skin friction relation (19) on p. 518 of the paper. Further, the power law velocity profile analysed in limit of vanishing viscosity by  Barenblatt \& Chorin [9] using physical variable $z = 1 - \ln \: y_+ / \ln \: Re$, is not the intermediate region in the pipe flow.

	The turbulent boundary layer by momentum integral method using power law velocity profile had been studied (see Schlichting [4]). The power law velocity profile was analysed by George \& Castillo [10], where predicted outer layer slip velocity is zero and for $R_\tau \ra \ity $ the data analysis predict $\al \ra 1/11$. Djenidi, Dubief \& Antonia [11] for boundary layer flow, following Barenblatt [2], proposed $\al = 1.5/ \ln  R_\de$ where $R_\de = U_\ity \de / \nu$.

\vs{2mm}
\noi {\bf 2. Analysis}\\
The turbulent boundary layer equations in standard notations [12] are
$$
u \fr{\pa u}{\pa x} + v \fr{\pa u}{\pa y} = -\fr{1}{\rho} \fr{\pa p}{\pa x} + \fr{1}{\rho} \fr{\pa \tau}{\pa y} + \nu \fr{\pa^2 u}{\pa y^2}, \hs{15mm} \fr{\pa u}{\pa x} + \fr{\pa v}{\pa y} = 0  \hs{10mm}
\eqno(2,3)
$$
$$
y = 0, \hs{5mm} u = v = \tau = 0; \hs{10mm} y/\de \ra \infty,  \hs{5mm}  u \ra U_\infty (x), \hs{5mm}  \tau \ra 0
\eqno(4)
$$
In the present work, the outer layer (the principal layer shown by Afzal[12]) is matched with inner layer.
$$
\mbox{Wake Layer:} \hs{4mm}u = U_\ity \:F'(X,Y), \hs{6mm} \tau = \rho\, u_\tau^2 \: T(X,Y), \hs{6mm} Y = y/\de, \hs{6mm} X = \int L^{-1} dx \hs{2mm}
\eqno(5)
$$
$$
T' + (a - \beta_c) F F'' - \beta_c(1-F'^2) + F'' F_x - F' {F'}_x = 0
\eqno(6)
$$
$$
\de/L = \ep^2, \hs{6mm} \ep = u_\tau/U_\ity. \hs{6mm}  \bt_c = - U_{\ity X}/U_\ity, \hs{10mm} a = \de_X/\de
$$
$$
u = U_\ity(X) \; U_0(X,Y) - u_\tau \; U_1(X,Y) \vs{4mm}
\eqno(7)
$$
$$
\mbox{ Wall Layer:} \hs{7mm} u = u_\tau \: u_+(X,y_+), \hs{10mm} \tau = \rho \, u_\tau^2 \; t(X,y_+),  \hs{10mm} y_+ = y u_\tau /\nu \hs{20mm}
\eqno(8)
$$
$$
u_+' + t = 1 - \La \;y_+ , \hs{10mm} \La = \nu \,p'/u_\tau^3 
$$
\noi {\bf Matching :} The matching of velocity profile $u$ by Millikan-Kolmogorov hypothesis, gives open functional equation and its solution, as shown by Afzal[12], leads to composite log laws. Another solution of functional equation  is  

\noi Wall power law :
$$
  u_+(X,y_+) = C \: y_+^\al, \hs{10mm} y_+ \ra \ity 
\eqno(9)
$$
\noi Wake power law :
$$
U_S/u_\tau - U_1(X,Y) = C_1 \: Y^\al, \hs{10mm} Y \ra 0
\eqno(10)
$$
$$
Y \pa U_0/\pa Y \ra 0 \hs{3mm} \mbox{as} \hs{3mm} Y \ra 0,   \hs {10mm} U_0(X,0) = b_S = U_S(X)/U_\ity (X)
$$
\noi Skin friction power law :
$$
U_s/u_\tau = C_1 + 2 E_S, \hs{10mm} C_1 = C \, R_\tau^\al,
\eqno(11)
$$
The lowest order wake layer equations under similarity
$$
U_0(X,Y) = g'(Y), \hs{10mm} \beta = m/(1+m)
\eqno(12)
$$
for $U_\ity(X)$ $\al$ $\de^m$ and Clauser's constant eddy viscosity $\al_c$ closure model, reduce to
$$
g''' + g g'' + \beta (1 - g'^2) = 0
\eqno(13)
$$
$$
g(0) = 0, \hs{10mm} g'(\ity ) = 1, \hs{10mm} g''(0) \int_0^\ity (1 - g') dY = \ep^2 / \al_c
\eqno(14)
$$
the Falkner-Skan equation with finite velocity slip $g'(0) = b_s$ and wall shear stress [12]. The uniformly valid solution for power law velocity profile above the sublayer is 
$$
u = U_\ity \left[ U_0(X,Y) - b_s \right] + u_\tau \left[ C y_+^\al + E_S \Omega (X,Y) \right]
\eqno(15)
$$
$$
(U_\ity - u)/U_\ity = b_s - U_0(X,Y) + (u_\tau / U_\ity)[C_1(-Y^\al +1) + E_S \{2 - \Omega (X,Y) \} ]
\eqno(16)
$$
$$
\Omega(X,Y) = [ - U_1(X,Y) + C_1 (1-Y^\al ) + 2 E_S ] / E_S
$$
where $\Omega$ is wake (power) function, $\Omega (X,0) = 0$ and $\Omega (X,1) = 2$. As $\al\ra 0$ for  $R_\tau \ra \ity$ then  $C_1/C=R_\tau^\al$ shall remain a constant of order unity, say $\ga$,  giving   $\al = \ga/\ln R_\tau$.

	For outer (Karman) defect layer solution is $U_0 (X,Y) = b_s = 1$ and with $E_S = E$ the uniformly valid power law solutions (15) and (16) for velocity  and  skin friction power law (11)  become
$$
u/u_\tau = C\: y_+^\al + E\: \Omega (X,Y)
\eqno(17)
$$
$$
(U_\ity - u)/u_\tau = C_1 (-Y^\al + 1) + E \: [2 - \Omega (X,Y)]
\eqno(18)
$$
$$
U_\ity / u_\tau = C \: R_\tau^\al + 2 E
\eqno(19)
$$
As $\al \ra 0$, $\ln \: y_+ \ra \ity $ the power law velocity with (1) is equivalent to log law velocity as
$$
C y_+^\al \ra \kappa^{-1} \ln \: y_+ + B, \hs{10mm} C_1 (-Y^\al +1) \ra -
\kappa^{-1} \ln \: Y, \hs{10mm} E \ra D
\eqno(20)
$$
\noi (where $2 D$ is intercept of defect log law). The expression (20) show that the relations (17)-(19) for $\al \ra 0$ (large Reynolds numbers) reduce to classical log laws for velocity profile and skin friction. Further, the wake (due to power law) function $\Omega (Y)$ reduces to Coles wake (due to log law)  function $W(Y)$.
\begin{figure}
\centering
    \includegraphics[width=0.8\textwidth]{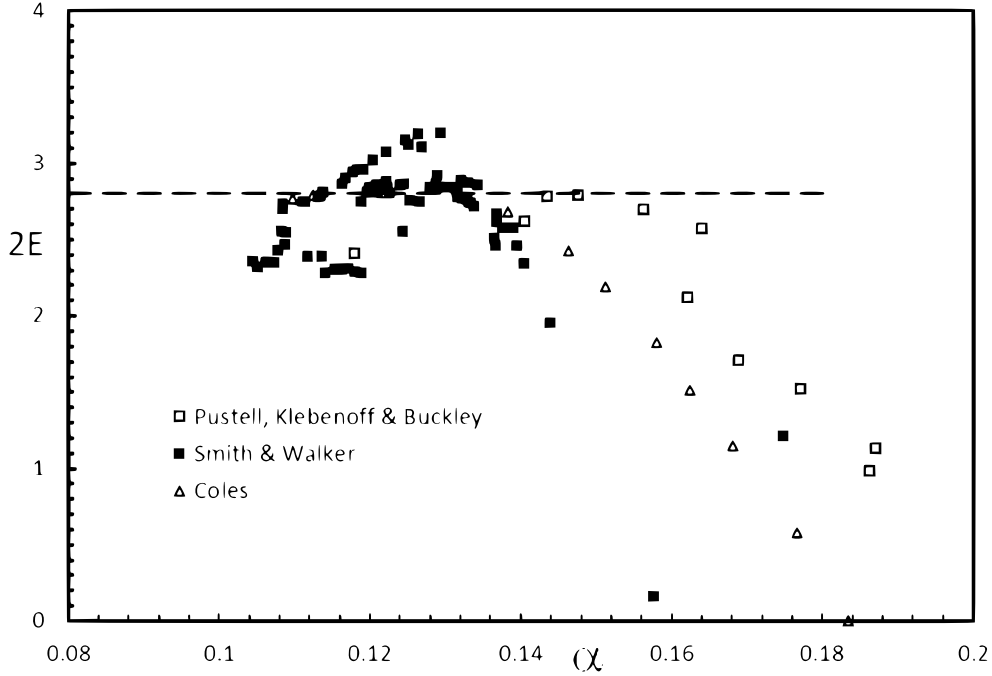}
\caption{The intercept 2E   against power law index $\al$ for  velocity defect layer (19).}
   \label{sol}
\end{figure}
%

\noi {\bf Result and Discussion}\\
For classical values $\kappa = 0.41$ and $B = 5$ [13], the power law constant $C$ from (1) is given by
$$
C \exp (1) = (0.41 \al)^{-1} + 5
\eqno(21)
$$
The outer layer power law intercept 2E has been estimated from relation (19) by employing 
the skin friction data (Coles [13], Smith \& Walker [14] and Purtell et al [15]) along with relations (1) and (21). Intercept 2E is displayed in Fig 1 against parameter $\al =  1/ \ln \: R_\tau$ (with $\ga=1$ for large Reynolds number). The data is in the range $0.1 < \al < 0.15 $, and within the data scatter it may  be represented by $2E = 4.75 D_0 (1-5.2 \al )$. The parameter $\la$ being the ratio of skin friction $C_f$ predicted by relation (19) to experimental value is slightly greater than unity for all values of Reynolds numbers. The predicted $\la$ displayed in Fig 2, based on relation (19) with $D_0 = 2$, compares well with experimental data of skin friction.
\begin{figure}
\centering
    \includegraphics[width=0.8\textwidth]{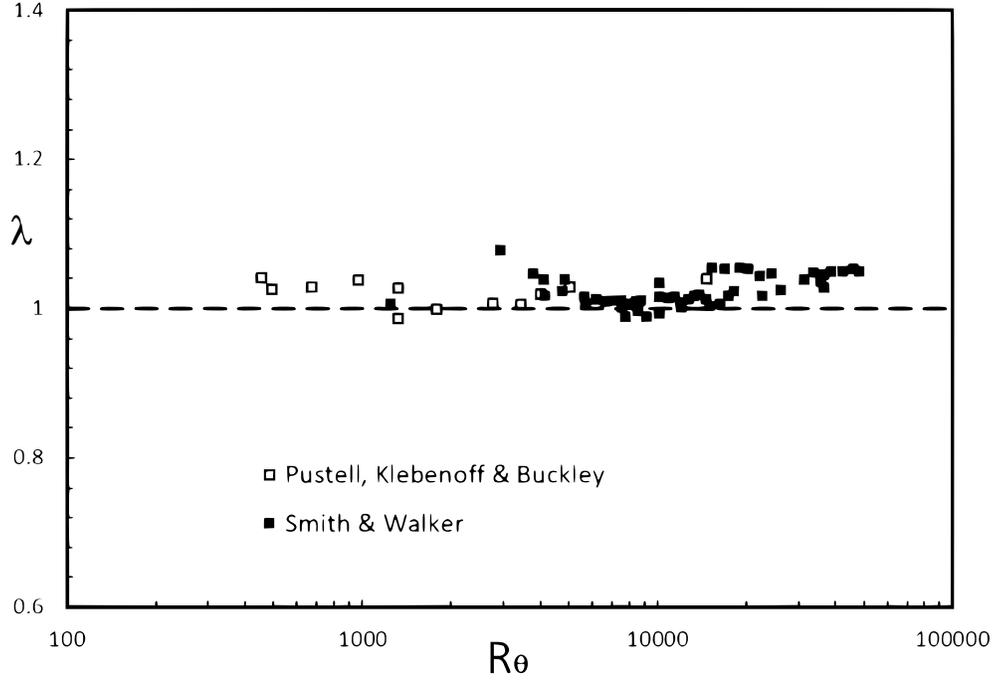}
\caption{Comparison of $\la$, the ratio of skin friction power law prediction to the experimental  value versus $R_\theta$ momentum Reynolds number  for constant pressure turbulent boundary layer data [14,15].}
   \label{sol2}
\end{figure}
%
%

\noi Acknowledgment : The author is thankful to CSIR for financial support.

\noi {\bf References}
\parskip3pt

\noi [1] Narasimha, R Invited Lecture in Proc. IUTAM Symposium (Ed: K Gersten), pp 5-16. \\ .\hs{7mm} Kluwer Academic Publishers Dordrecht / Boston / London (1996).

\noi  [2] Barenblatt, GI  J. Fluid Mech. 248, 513-520 (1993).

\noi  [3] Prandtl, L In Aerodynamic Theory. (Ed: Durand) Vol. 3, pp. 132-3, (1935).

\noi  [4] Schlichting, H Boundary Layer Theory. McGraw Hill New York, 1968.

\noi  [5] Kailasnath, P Ph.D. Thesis, Mason Lab., Yale University, 1993.
\newpage
\noi  [6] Afzal, N Connection between power and log laws for turbulent flow. International \\ .\hs{7mm} Conference on Functional Analysis \& Applications (Ed: A. H. Siddiqui), pp 7-8, held  \\ .\hs{7mm} on 16-19 Dec. 1996, Mathematics Department, Aligarh Muslim University, Aligarh.

\noi  [7] Zagarola MV, Perry AE \& Smits AJ Phy. Fluids 9, 2094-2100 (1997).

\noi  [8] Afzal, N \& Narasimha, R JNCASR Bangalore, Rept 1997 (To appear).

\noi  [9] Barenblatt, GI \& Chorin, AJ Comm. Pure \& Appl. Maths 50, 381-398 (1997).

\noi  [10] George, W \& Castillo, L In Near Wall Turbulent Flow (Ed: RMC So, CG Speziale \& \\ .\hs{7mm} BE Launder), pp 901-910. Elsevier Science Publishers BV (1993).

\noi  [11] Djenidi L, Dubief R \& Antonia RA Exp. Fluids 22, 348-350 (1997).

\noi  [12] Afzal, N Invited Lecturer in Proc. IUTAM Symposium (Ed: K Gersten), pp 95-118. Kluwer Academic Publishers Dordrecht / Boston / London (1996).

\noi  [13] Coles, D Rand Corporation Report R-403-PR (1962).

\noi  [14] Smith, DW \& Walker, JH NACA Tech. Note 4231 (1959).

\noi  [15] Purtell, LP, Klebanoff, PS \& Buckley FT, Phy. Fluids 24, 802-811 (1981)

\end{document}